\documentclass[twoside,11pt]{article}

\usepackage{amsmath}
\usepackage{amssymb}
\usepackage{bm}
\usepackage{dsfont}
\usepackage{graphicx}
\usepackage{esint}
\usepackage[margin=1in]{geometry}
\usepackage[biblabel=brackets,style=phys,eprint=true]{biblatex}
\usepackage{xcolor}
\usepackage{hyperref}
\usepackage{mathtools}
\usepackage{authblk}
\usepackage{subcaption}
\usepackage{array}
\usepackage{sidecap}
\usepackage[titletoc,toc,page]{appendix}
\usepackage{etoolbox}

\sidecaptionvpos{figure}{c}

\title{Neural Networks as Spin Models: From Glass to Hidden Order Through Training}
\author[1]{Richard Barney}
\author[1,2]{Michael Winer}
\author[1]{Victor Galitski}
\affil[1]{Joint Quantum Institute, Department of Physics, University of Maryland, College Park, Maryland, USA}
\affil[2]{Institute for Advanced Study, Princeton University, Princeton, New Jersey, USA}
\date{}

\begin{document}
\maketitle
\begin{abstract}
We explore a one-to-one correspondence between a neural network (NN) and a statistical mechanical spin model where neurons are mapped to Ising spins and weights to spin-spin couplings. The process of training an NN produces a family of spin Hamiltonians parameterized by training time. We study the magnetic phases and the melting transition temperature as training progresses. First, we prove analytically that the common initial  state before training---an NN with independent random weights---maps to a layered version of the classical Sherrington-Kirkpatrick spin glass exhibiting a replica symmetry breaking. The spin-glass-to-paramagnet transition temperature is calculated. Further, we use the Thouless-Anderson-Palmer (TAP) equations---a theoretical technique to analyze the landscape of energy minima of random systems---to  determine the evolution of the magnetic phases on two types of NNs (one with continuous and one with binarized activations) trained on the MNIST dataset. The two NN types give rise to  similar results, showing a quick destruction of the spin glass and the appearance of a phase with a hidden order, whose melting transition temperature $T_c$ grows as a power law in training time. We also discuss the properties of the spectrum of the spin system's bond matrix in the context of rich vs. lazy learning. We suggest that this statistical mechanical view of NNs provides a useful unifying perspective on the training process, which can be viewed as selecting and strengthening a symmetry-broken state associated with the training task. 
\end{abstract}

\tableofcontents

\section{Introduction}

The synergy between machine learning and statistical mechanics has been a long-standing topic of research~\cite{amit1985,dotsenko1994,nakanishi1997,barra2008,barra2010,barra2012,choromanska2015,baity-jesi2018}. In particular, there is a natural relation between spin systems and neural networks (NNs), including multilayer perceptrons~\cite{almeida1996}, restricted Boltzmann machines~\cite{zhang2018}, and Hopfield networks~\cite{hopfield1982,wen2009,krotov2023}, where neurons correspond to the spin degrees of freedom, the biases to magnetic fields, and the weights to the spin-spin couplings. For a random realization of weights between neurons, spin glass physics naturally becomes relevant, along with the methods of studying glasses such as replica symmetry breaking~\cite{parisi1979,parisi1980,parisi1983,mezard1986,Fischer_Hertz_1991} and the Thouless-Anderson-Palmer (TAP) equations~\cite{thouless1977,plefka1982,Fischer_Hertz_1991}. These methods are potentially useful in transplanting intuition from statistical mechanics (phase transitions, universality, emergent behavior, etc.) to the more practical field of NNs, with the potential to elucidate the underlying mechanisms for when and how they work in accomplishing various tasks. 

Another motivation to study this machine learning/statistical mechanics correspondence is the recent progress in the field of neuromorphic computing~\cite{markovi2020,schuman2022}, which employs physical analogue networks rather than algorithmic networks. While still at the experimental state of development, there has been notable recent success in building such NNs with ultracold atoms in cavity QED systems. Unlike feed-forward NNs, there is no unidirectional layered structure in such systems (nor in biological NNs), yet they share similarities with current commercial machine learning platforms and show potential for performing useful tasks. While most work so far has focused on experimentally probing classical concepts (associative memory~\cite{marsh2021}, replica symmetry breaking~\cite{kroeze2023,marsh2024}, etc.) with these systems, they include quantum components from the outset. Hence putting together an underlying theoretical framework to include both classical and quantum NNs is a topic of great interest.

This paper studies two types of NNs with distinct training processes for a classification task from the perspective of statistical mechanics, in particular using the TAP equations. These are conventionally used in the studies of random spin systems to quantitatively characterize their energy landscapes and identify phases and phase transitions. We argue that the TAP equations also provide both a powerful tool to explore traditional NNs and a bridge to their quantum counterpart. The latter is left for  future work, while here we study the training process of an NN viewed as a family of classical spin Hamiltonians and explore their phases as a function of ``time''---the epoch of the training process. 

Specifically, we consider two types of feed-forward NNs, one with step and one with ReLU activations. We train the NNs on an image classification task: recognition of images of handwritten digits from $0$ to $9$ with the MNIST dataset~\cite{Lecun1998}. At each step of training, which we parameterize by training time $t$ measured in epochs, we have a set of weights $J_{ij}(t)$. We associate each such instance with a classical Ising spin Hamiltonian, where the weights are the couplings between the spins, and explore the flow of its energy landscape and phases using the TAP equations.

\begin{figure}[h!]
    \centering
    \includegraphics[width=0.6\linewidth]{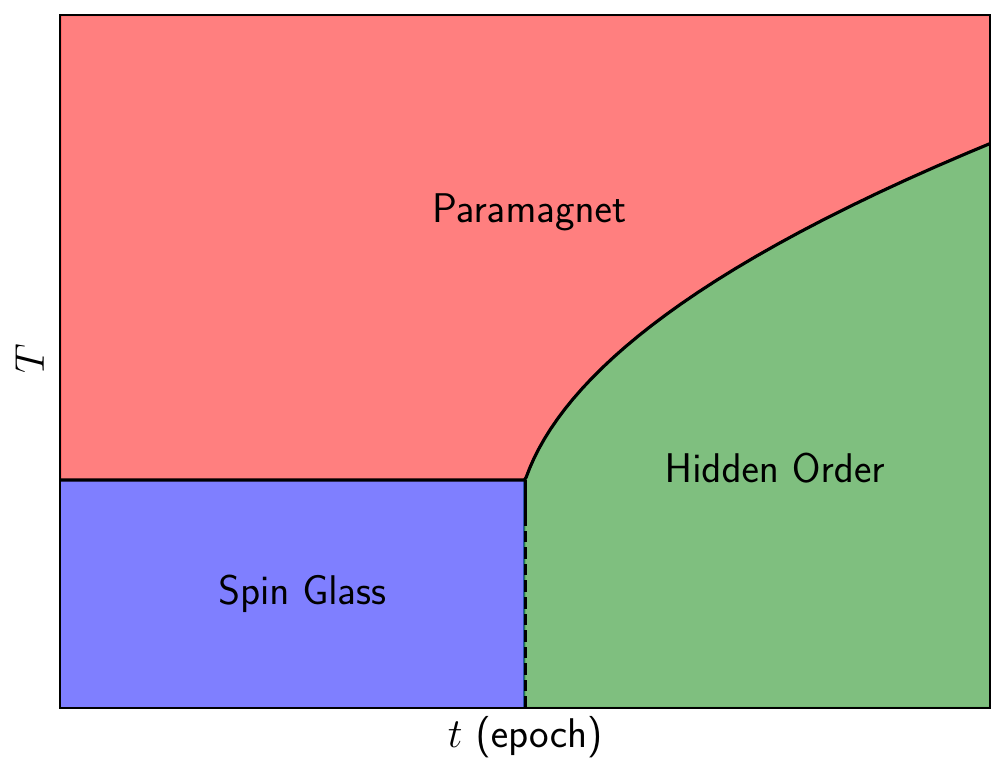}
    \caption{A schematic temperature vs. training epoch phase diagram for a multi-layer spin model with bonds determined by neural network training for temperatures not significantly below the melting temperature. The critical temperature grows as a power law with training time.}
    \label{fig:phase_diagram}
\end{figure}

The conventional starting point of training, at $t=0$, is a random realization of couplings. We show analytically that its spin counterpart exhibits a spin glass transition with replica symmetry breaking and calculate the melting temperature $T_c$ of the glass. Throughout the training process we numerically calculate $T_c$ by determining the temperature at which nontrivial solutions to the TAP equations appear. We note distinct regimes of power-law scaling of the transition temperature as a function of training, $T_c(t) \propto t^\alpha$, where $\alpha$ appears to depend on the rate at which the NN is learning. We also consider the number of TAP solutions that appear at the transition and observe a rapid flow of the spin Hamiltonian from the initial glassy Hamiltonian into one with a $\mathbb Z_2$ symmetry breaking phase transition. We note a further enhancement of the hidden order, which manifests itself in the increase in $T_c$ with training.  We speculate that the hidden order encodes the classification task, where the outcome---the output layer---is characterized by a set of parameters much smaller than the width of a layer (512 neurons) which encodes a hidden order parameter. Schematically, we observe a phase diagram as shown in Fig.~\ref{fig:phase_diagram} for temperatures not much below $T_c$. We determine that these changes can be primarily attributed to the dynamics of the spectrum of the bond matrix $J(t)$. We also comment on the qualitative similarities between the final bond matrix spectrum and the spectra of the weight matrices connecting adjacent layers of NNs in the rich learning regime, as observed in previous works~\cite{matthias2022,martin2021,martin2021_predicting,wang2024}.

The remainder of this work is organized as follows. In Section~\ref{sec:init} we analyze the spin system corresponding to our NNs before training. In doing so we introduce the multi-layer Sherrington-Kirkpatrick spin glass model and analytically identify the critical temperature at which replica symmetry breaking occurs. We also introduce the TAP equations and describe how they may be used to find the critical temperature for a single instance of a multi-layer spin model with bonds determined by some training process. In Section~\ref{sec:training} we describe the training procedures for our NNs, discuss their performance, determine how the critical temperature evolves as the training progresses, and discuss the changing nature of the melting transition. We then examine the evolution of the spectrum of the bond matrix, which is the primary driver of the previously discussed changes, and connect to previous work on rich vs. lazy learning. We finish by summarizing our conclusions and future directions in Section~\ref{sec:conclusion}.

\section{The initial state of the neural network}\label{sec:init}

Before any training occurs the weights in our NNs are independently and normally distributed in such as way as to preserve the magnitude of the variance of the activations in the forward pass, as in Kaiming normal initialization~\cite{kaiming2015}. This means that the weights fed into the $k^\text{th}$ layer are independently drawn from the normal distribution $\mathcal N(0,C_k/N_{k-1})$, where $C_k$ is $O(1)$ and $N_k$ is the number of neurons in the $k^\text{th}$ layer. For the sake of generality we allow $C_k$ to be different between layers for now.

\begin{figure}[h!]
    \centering
    \includegraphics[width=0.7\linewidth]{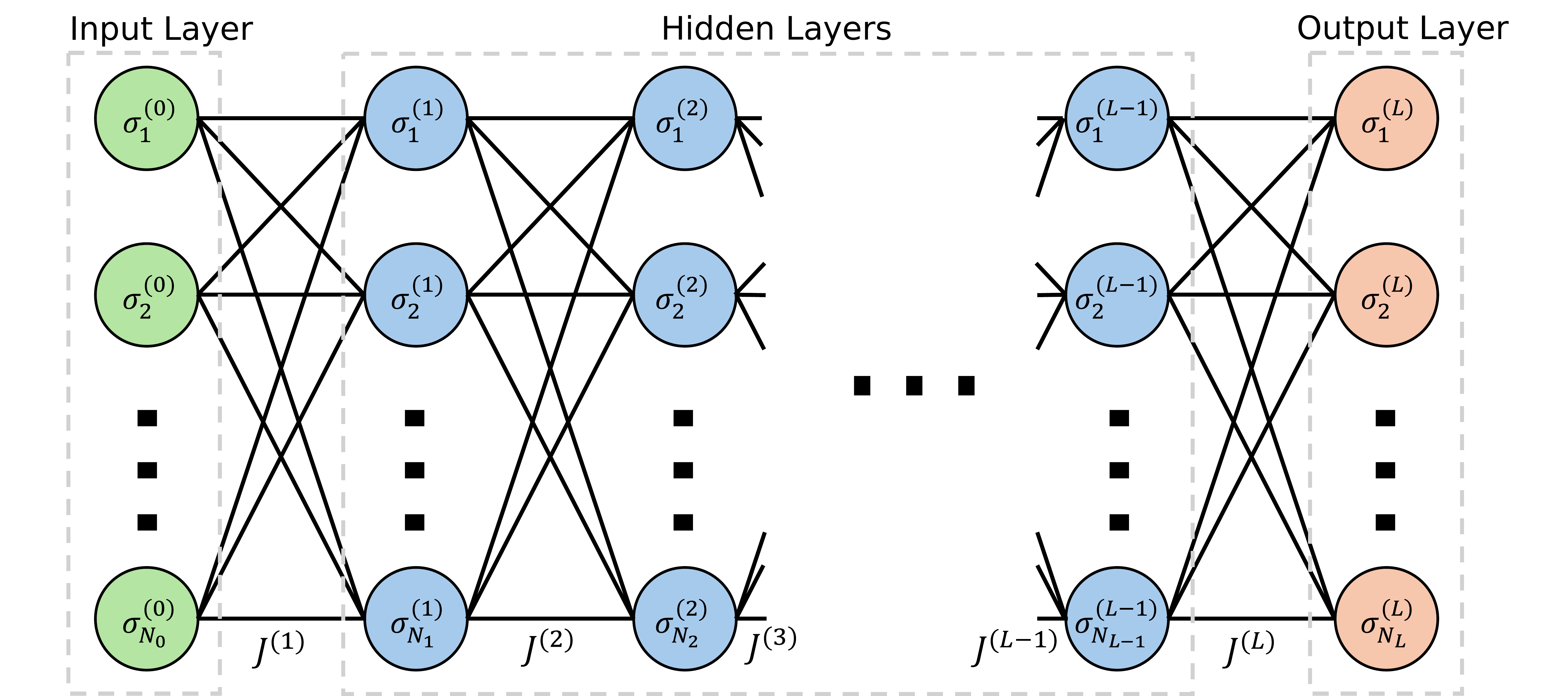}
    \caption{Schematic representation of a neural network with $L+1$ layers. The $k^\text{th}$ layer contains $N_k$ neurons. $J^{(k)}$ is the weight matrix connecting layer $k-1$ to layer $k$.}
    \label{fig:NN_fig}
\end{figure}

Consider an NN with $L+1$ layers labelled $0,\dots,L$, as shown in Fig.~\ref{fig:NN_fig}. By mapping the neurons in the NN onto Ising spins, which can take only the values $\pm 1$, and using the weights as interactions between the spins, we obtain the multi-layer Sherrington-Kirkpatrick (MSK) model~\cite{agliari2021} defined by the Hamiltonian
\begin{equation}
    H_{\text{MSK}}(\sigma)=-\sum_{k=1}^L\sum_{i,j=1}^{N_{k-1},N_k}J_{ij}^{(k)}\sigma_i^{(k-1)}\sigma_j^{(k)},
\end{equation}
where each $J_{ij}^{(k)}$ is independently drawn from $\mathcal N(0,C_k/N_{k-1})$. For large $N_k$ each spin has a high coordination number, so we can expect mean field theory to have good predictive power. MSK is a slight modification of the standard well-known Sherrington-Kirkpatrick (SK) model~\cite{sherrington1975}, in which independent and identically distributed interactions exist between every pair of spins in the system, having no layered structure. The MSK model can also be viewed as a specific case of a multi-species spin glass~\cite{barra2015}. 

Later it will become convenient to label the spins with a single index such that
\begin{equation}
    \sigma_i^{(k)}=\sigma_{i+\sum_{\ell=0}^{k-1}N_\ell}.
\end{equation}
With this we can write the MSK Hamiltonian in the simpler form
\begin{gather}
    H_\text{MSK}(\sigma)=-\frac 1 2\sum_{i,j=1}^{N}J_{ij}\sigma_i\sigma_j,\\
    J=\begin{pmatrix}
        0 & J^{(1)} & 0 & \cdots & 0\\
        J^{(1)T} & 0 & J^{(2)} & \ddots & \vdots\\
        0 & J^{(2)T} & \ddots & \ddots & 0\\
        \vdots & \ddots & \ddots & 0 & J^{(L)}\\
        0 & \cdots & 0 & J^{(L)T} & 0
    \end{pmatrix},
\end{gather}
where $N=\sum_{k=0}^LN_k$.

\subsection{Transition temperature}\label{ssec:analytic}

For sufficiently high temperatures we reasonably expect the MSK model to be in the paramagnetic phase. That is, the system will behave ergodically. We wish to find the critical temperature at which a non-ergodic phase appears. We expect that, as with the SK model, this low-temperature phase will be a spin glass. This glassy phase is characterized by replica symmetry breaking~\cite{parisi1979,parisi1980,parisi1983,mezard1986,Fischer_Hertz_1991}. If a large number of exact replicas of the system were made with identical realizations of the disorder and all were rapidly cooled, the replicas would settle into a large number of different neighborhoods around local energy minima.

We could make use of the replica ``trick"~\cite{edwards1975,mezard1986} to find the disorder averaged expectation of some variable $A$ as
\begin{equation}
    \langle \mathbb E(A)\rangle=\left\langle\frac{1}{Z}\sum_{\sigma_1} A(\sigma_1)e^{-\beta H(\sigma_1)}\right\rangle=\lim_{m\rightarrow 0}\left\langle Z^{m-1}\sum_{\sigma_1} A(\sigma_1)e^{-\beta H(\sigma_1)}\right\rangle=\lim_{m\rightarrow 0}\sum_{\{\sigma_\alpha\}_{\alpha=1}^m} A(\sigma_1)\langle Z^m\rangle,
\end{equation}
where $\beta$ is the inverse temperature, $Z$ is the partition function, and the replicated partition function is
\begin{equation}
    Z^m=\exp\left[-\beta\sum_{\alpha=1}^mH(\sigma_\alpha)\right].
\end{equation}
We are therefore interested in the disorder averaged replicated partition function for small $m$. We search for the temperature at which the replica symmetric solution becomes unstable, indicating the breaking of replica symmetry and the appearance of the glassy phase.

For each layer we define the $m\times m$ overlap matrix $Q_{\alpha\beta}^{(k)}=\frac{1}{N_k}\sum_{i=1}^{N_k}\sigma_{i\alpha}^{(k)}\sigma_{i\beta}^{(k)}$. We enforce these conditions with the Lagrange multipliers $\Lambda_{\alpha\beta}^{(k)}$. We then find after disorder averaging that
\begin{gather}
    \langle Z^m\rangle=\int\mathcal DQ\mathcal D\Lambda~e^S,\\
    S=\frac 12\beta^2\sum_{k=1}^L N_k\sum_{\alpha\beta} C_k Q_{\alpha\beta}^{(k-1)}Q_{\alpha\beta}^{(k)}+\sum_{k=0}^L N_k\left[\sum_{\alpha\beta}\frac 12\Lambda_{\alpha\beta}^{(k)}Q_{\alpha\beta}^{(k)}+\log\sum_{\{\sigma_\alpha\}}\exp \left(-\frac 12\sum_{\alpha\beta}\sigma_\alpha\Lambda_{\alpha\beta}^{(k)}\sigma_\beta\right)\right].
\end{gather}
We now make the replica symmetric ansatz
\begin{gather}
    Q_{\alpha\beta}^{(k)}=q^{(k)}+\delta_{\alpha\beta}(p^{(k)}-q^{(k)}),\\
    \Lambda_{\alpha\beta}^{(k)}=\lambda^{(k)}+\delta_{\alpha\beta}(\mu^{(k)}-\lambda^{(k)})
\end{gather}
and expand the action to second order in $q$ and $\lambda$. This yields
\begin{equation}
    S/m\approx \frac{\beta^2}{2}\sum_{k=1}^LN_kC_k(p^{(k-1)}p^{(k)}-q^{(k-1)}q^{(k)})+\frac 12\sum_{k=0}^LN_k\left(\mu^{(k)}(p^{(k)}-1)-\lambda^{(k)}q^{(k)}-\frac 12(\lambda^{(k)})^2\right),
\end{equation}
ignoring higher order terms in $m$ and constant terms. We will evaluate the integrals over $Q$ and $\Lambda$ by the saddle point method. By differentiating the equation for the action above in terms of $\mu^{(k)}$ and $\lambda^{(k)}$ we find the saddle point equations
\begin{gather}
    p^{(k)}=1\\
    \lambda^{(k)}=-q^{(k)}.
\end{gather}
With these we can write the action as
\begin{equation}
    S/m\approx-\frac{\beta^2}{2}\sum_{k=1}^LN_kC_kq^{(k-1)}q^{(k)}+\frac 14\sum_{k=0}^LN_k(q^{(k)})^2.
\end{equation}

The Hessian matrix of the action is then proportional to the tridiagonal matrix
\begin{equation}\label{eq:Hessian}
    \mathbf H=\begin{pmatrix}
        N_0 & -\beta^2N_1C_1 & 0 & \cdots & 0\\
        -\beta^2N_1C_1 & N_1 & -\beta^2N_2C_2 & \ddots & \vdots\\
        0 & -\beta^2N_2C_2 & N_2 & \ddots & 0\\
        \vdots & \ddots & \ddots & \ddots & -\beta N_L C_L\\
        0 & \cdots & 0 & -\beta N_L C_L & N_L
    \end{pmatrix}.
\end{equation}
At sufficiently high temperatures this matrix is positive definite and the replica symmetric solution will be stable. This indicates that the system is in the paramagnetic phase. When the temperature decreases to the point that the Hessian is no longer positive definite, the replica symmetric solution is no longer stable and replica symmetry breaking occurs. This indicates the transition to the glassy phase.


For the simple case in which each layer is identical, letting all $N_k=n$ and $C_k=C$, the Hessian becomes a tridiagonal Toeplitz matrix with the eigenvalues
\begin{equation}
    n\left[1-2\beta^2C\cos\left(\frac{(k+1)\pi}{L+2}\right)\right],\quad k=0,\dots,L.
\end{equation}
The least eigenvalue vanishes at the transition temperature
\begin{equation}\label{eq:an_trans_temp}
    T_c=\left[2C\cos\left(\frac{\pi}{L+2}\right)\right]^{1/2}.
\end{equation}
If the NN is very deep with large $L$, the transition temperature approaches
\begin{equation}
    T_c=\sqrt{2C}.
\end{equation}

\subsection{Thouless-Anderson-Palmer Equations}
So far our examination of the multi-layer SK model has made use of disorder averaging. That is, we have been examining an ensemble of systems with independent randomly distributed bonds. However, when we are examining systems in which the bonds are determined by some training procedure, we no longer have such an ensemble to average over. We can instead make use of the Thouless-Anderson-Palmer (TAP) approach~\cite{thouless1977,plefka1982}, which allows us to analyze an energy landscape for a single instance.

The mean field free energy for a system with pairwise interactions between spins is
\begin{multline}
    F=-\frac{1}{2}\sum_{ij}J_{ij}m_im_j-\frac 14\beta\sum_{ij}J_{ij}^2(1-m_i^2)(1-m_j^2)\\
    +\frac 12T\sum_i\left\{(1+m_i)\log\left[\frac 12(1+m_i)\right]+(1-m_i)\log\left[\frac 12(1-m_i)\right]\right\},  
\end{multline}
where $m_i=\langle\sigma_i\rangle$. This is the Curie-Weiss free energy~\cite{kac1969} with the addition of the second term on the right, known as the Onsager term~\cite{onsager1936}. This term describes the contribution to the effective field felt by a spin due to the spin itself. Since the effective field should be determined only by the other spins, this contribution is subtracted. By minimizing with respect to the $m_i$ we obtain the TAP equations
\begin{equation}
    \tanh^{-1}m_i=\beta\sum_j J_{ij}m_j-\beta^2\sum_j J_{ij}^2(1-m_j^2)m_i.
\end{equation}
The solutions to these equations are the locations of fixed points of the mean field free energy.

At temperatures not much lower than the transition temperature we can take each $m_i$ to be small, allowing us to linearize the TAP equations to
\begin{equation}
    m_i=\beta\sum_j J_{ij}m_j-\beta^2\sum_jJ_{ij}^2m_i.
\end{equation}
We can also express these as the matrix equation
\begin{gather}
    (I_N-M)\mathbf m=0,\\
    M_{ij}=\beta J_{ij}-\delta_{ij}\beta^2\sum_\ell J_{i\ell}^2.\label{eq:M_gen}
\end{gather}
Above the transition temperature the only solution is the trivial $\mathbf m=0$, indicating the paramagnetic phase. At the transition temperature $I_N-M$ will cease to be positive definite as at least one eigenvalue vanishes and at least one nontrivial solution appears. Due to the layered structure of our model, $M$ becomes the block tridiagonal matrix
\begin{gather}
    M=\beta\begin{pmatrix}
        R^{(0)} & J^{(1)} & 0 & \cdots & 0\\
        J^{(1)T} & R^{(1)} & J^{(2)} & \ddots & \vdots\\
        0 & J^{(2)T} & R^{(2)} & \ddots & 0\\
        \vdots & \ddots & \ddots & \ddots & J^{(L)}\\
        0 & \cdots & 0 & J^{(L)T} & R^{(L)}
    \end{pmatrix},\label{eq:M_block}\\
    R_{ij}^{(k)}=-\delta_{ij}\beta\begin{cases}
        \sum_{\ell=1}^{N_1}(J_{i\ell}^{(1)})^2, & k=0\\
        \sum_{\ell=1}^{N_{k-1}}(J_{\ell i}^{(k)})^2+\sum_{\ell=1}^{N_{k+1}}(J_{i\ell}^{(k+1)})^2, & 1\leq k\leq L-1\\
        \sum_{\ell=1}^{N_{L-1}}(J_{\ell i}^{(L)})^2, & k=L
    \end{cases}\label{eq:R}.
\end{gather}

For not-too-large systems the transition temperature can be found numerically by exact diagonalization, which is the approach we take in this work. For deep NNs with a large number of layers a more efficient approach is to consider that the block tridiagonal matrix $I_N-M$ will be positive definite as long as all $\chi_k$ satisfying the recurrence relation
\begin{equation}
    \chi_k=\begin{cases}
        \beta R^{(0)}-I_{N_0}, & k=0\\
        \beta R^{(k)}-I_{N_k}-\beta^2J^{(k)T}\chi_{k-1}^{-1}J^{(k)}, & 1\leq k\leq L
    \end{cases}
\end{equation}
are positive definite~\cite{el-mikkawy2003}. The determinant is then 
\begin{equation}
    \det I_N-M=\prod_{i=0}^L\det\chi_k.
\end{equation}
The transition temperature can then be determined by finding the value of $\beta$ for which the determinant vanishes, meaning that at least one of the $\chi_k$ has a vanishing eigenvalue.

\begin{figure}[h!]
    \centering
    \includegraphics[width=0.7\textwidth]{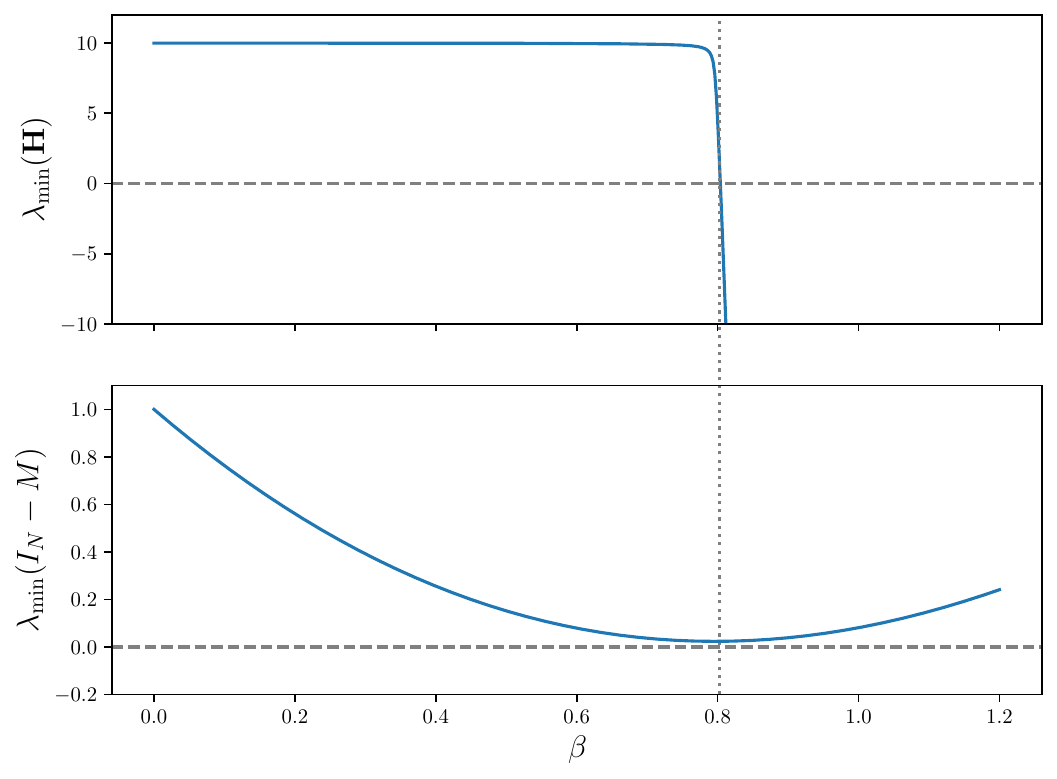}
    \caption{(Top) The least eigenvalue of the Hessian of the action [Eq.~(\ref{eq:Hessian})] for the ensemble describing our neural networks before training. (Bottom) The least eigenvalue of $I_N-M$ for a particular system in that ensemble. Both vanish or nearly vanish at the same value of $\beta$.}
    \label{fig:compare_least}
\end{figure}

For systems with sufficiently large layers we expect the result from the TAP equations for a single instance to approximate our analytical result for the ensemble. In Fig.~\ref{fig:compare_least} we show the least eigenvalues of $I_N-M$ for a single system in the ensemble and the Hessian $\mathbf H$ of the action for the entire ensemble. We see that the least eigenvalues of both do indeed vanish (or nearly vanish due to finite size effects) at the same temperature. As the system moves to the thermodynamic limit we expect the least eigenvalue of $I_N-M$ to vanish at the single point of its minimum as a function of $\beta$. This indicates that nontrivial TAP solutions appear at the same temperature that replica symmetry breaking occurs in the MSK model.

\section{Evolution with training}\label{sec:training}
\subsection{The training programs}

We trained multiple NNs on the MNIST dataset~\cite{Lecun1998}. This dataset is comprised of a $28\times 28$ grayscale images of handwritten digits 0-9. The training set contains 60,000 images while the test set contains 10,000. The classification task is to label each image with the correct digit. Each of our NNs had three hidden layers containing 512 nodes each.

We analyzed the results of two distinct training programs. The first was designed to closely follow the analogy between neurons in the NN and Ising spins. This was done by training a partially binarized NN. Each neuron in the NN can only take on the values $\pm 1$, but the weights between neurons in adjacent layers are still continuous variables. This is in contrast to NNs which binarize only the weights~\cite{courbariaux2016binaryconnect} or fully binarized NNs~\cite{courbariaux2016} which binarize both activations and weights. The binarization is enforced by applying the forward activation function
\begin{equation}
    \phi_\text{forward}(x)=\text{step}(x)=\begin{cases}
        -1, & x<0\\
        1, & x>0
    \end{cases}.
\end{equation}
between each layer of the NN, as well as on the input data. We will refer to this as the binarized NN.

In order to train the binarized NN effectively, we use a modified version of the straight through estimator~\cite{bengio2013} when calculating the gradients. This amounts to using the hard hyperbolic tangent (htanh) activation function
\begin{equation}
    \phi_\text{back}(x)=\text{htanh}(x)=\begin{cases}
        -1, & x<-1\\
        x, & -1<x<1\\
        1, & x>1
    \end{cases}
\end{equation}
during the backpropagation step. It has been shown that using such a modified STE in a fully binarized NN allows for near state-of-the-art performance while significantly reducing the computation resources necessary for a forward pass \cite{courbariaux2016}. The loss function for our NN is the $L_2$ multi-class hinge loss~\cite{boser1992,tang2013}. The NN is then trained using stochastic gradient descent (SGD) with momentum.

The second training program is more typical, with neurons able to take on continuous activations. The activation function $\phi(x)$ used in the hidden layers is the popularly used rectified linear unit (ReLU)
\begin{equation}
    \phi(x)=\text{ReLU}(x)=\begin{cases}
        0, & x\leq 0\\
        x, & x>0
    \end{cases}.
\end{equation}
We will refer to this NN as the standard NN. Once again the loss function is the multi-class hinge loss and training is done by SGD with momentum. In this case the final layer is an $L_2$ support vector machine.

\begin{figure}[h!]
    \centering
    \begin{subfigure}{0.4\textwidth}
        \caption{Binarized}
        \includegraphics[width=\linewidth]{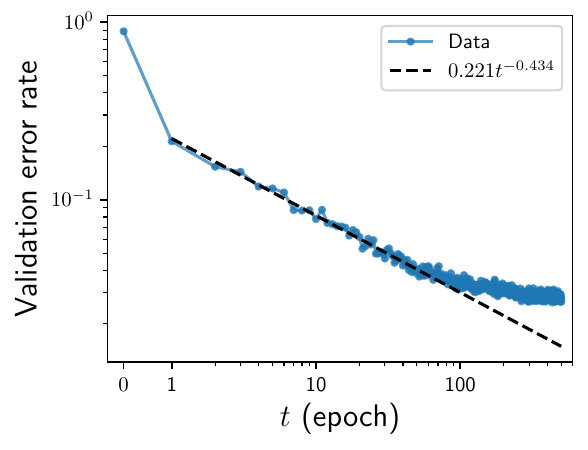}
    \end{subfigure}
    \begin{subfigure}{0.4\textwidth}
        \caption{Standard}
        \includegraphics[width=\linewidth]{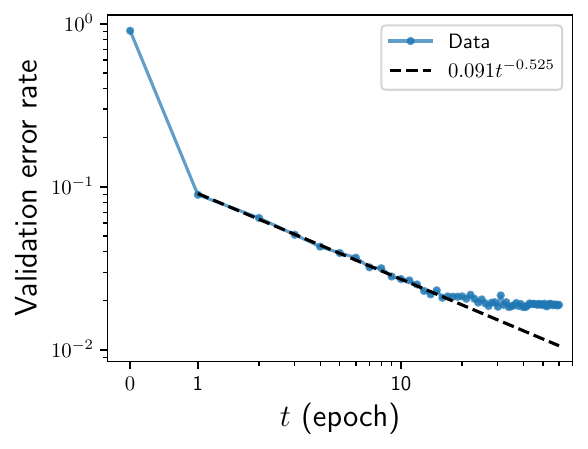}
    \end{subfigure}
    \caption{The validation error rates as training progresses. All axes are log scaled excepting the linear scaling between 0 and 1 on the horizontal axis. The learning rate for these NNs is $10^{-4}$. Data is collected at the end of each epoch.}
    \label{fig:err_rate}
\end{figure}

Table~\ref{tab:params} summarizes the relevant training parameters for both training procedures and the resulting validation error rates. As can be seen in Fig.~\ref{fig:err_rate}, for the larger learning rate, after 60 training epochs the validation error rate of the standard NN reaches approximately 0.019 while the binarized NN reaches approximately 0.027 after 500 epochs. Note that the error rates of both NNs initally decay as power laws but then eventually plateau.

\begin{table}[h!]
    \centering
    \begin{tabular}{|c||c|c|c|c|}
        \hline
        NN type & \multicolumn{2}{c|}{Binarized} & \multicolumn{2}{c|}{Standard}\\
        \hline
        Forward activation & \multicolumn{2}{c|}{step} & \multicolumn{2}{c|}{ReLU}\\
        \hline
        Backward activation & \multicolumn{2}{c|}{htanh} & \multicolumn{2}{c|}{ReLU}\\
        \hline
        Learning rate &  $10^{-4}$ & $10^{-6}$ & $10^{-4}$ & $10^{-6}$\\
        \hline
        Training epochs & 500 & 960 & 60 & 960\\
        \hline
        Validation error rate & 0.0266 & 0.0742 & 0.0189 & 0.0267\\
        \hline
        Momentum & \multicolumn{4}{c|}{0.9}\\
        \hline
        Batch size & \multicolumn{4}{c|}{32}\\
        \hline
        \# hidden layers & \multicolumn{4}{c|}{3}\\
        \hline
        Hidden layer size & \multicolumn{4}{c|}{512}\\
        \hline
        Loss function & \multicolumn{4}{c|}{$L_2$ multi-class hinge}\\
        \hline
        Margin & \multicolumn{4}{c|}{10}\\
        \hline
    \end{tabular}
    \caption{The neural network training details.}
    \label{tab:params}
\end{table}

\subsection{Evolution of the critical temperature}
\begin{figure}[h!]
    \centering
    \begin{subfigure}{0.4\textwidth}
        \caption{Binarized}
        \includegraphics[width=\linewidth]{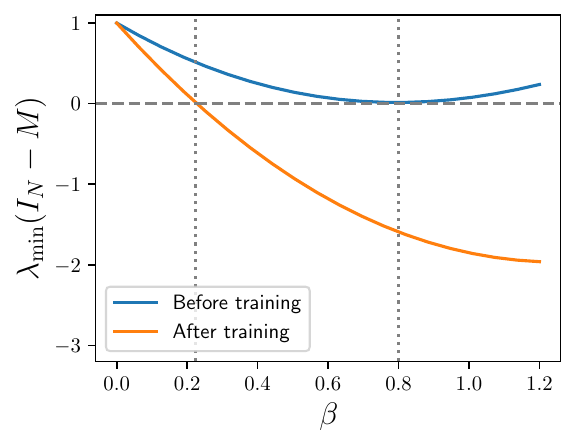}
    \end{subfigure}
    \begin{subfigure}{0.4\textwidth}
        \caption{Standard}
        \includegraphics[width=\linewidth]{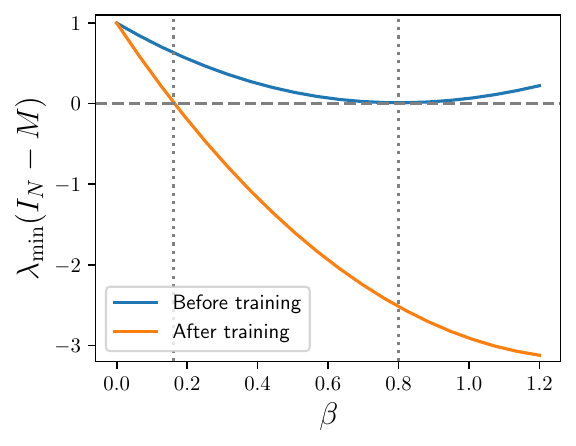}
    \end{subfigure}
    \caption{The least eigenvalue of $I_N-M$ as a function of inverse temperature $\beta$ before and after training the neural networks. The learning rate for these neural networks is $10^{-4}$.}
    \label{fig:min_eig_evo}
\end{figure}

The weights obtained throughout training allow us to define a family of Hamiltonians for a multi-layer spin system parameterized by the training time $t$:
\begin{equation}
    H(\sigma,t)=-\frac 12\sum_{i,j=1}^NJ_{ij}(t)\sigma_i\sigma_j.
\end{equation}
A natural question is how the critical temperature of this system changes as training progresses. To accomplish this we determine the least eigenvalue of $I_N-M$ over a range of $\beta$ by exact diagonalization, where $M$ is given in Eqs.~(\ref{eq:M_block})-(\ref{eq:R}). Fig.~\ref{fig:min_eig_evo} shows the least eigenvalue vs. $\beta$ before and after training for our two types of NNs. Before training the least eigenvalue vanishes only at its minimum (or nearly does so). As training progresses, the minimum drops and the least eigenvalue first vanishes at a smaller $\beta$. We observe that the least eigenvalue curve drops significantly lower for the standard NN than for the binarized NN. This is likely due to more effective training in the standard NN since the activations are not constrained to be binary and errors are not introduced by having different forward and backward activation functions.

\begin{figure}[h!]
    \centering
    \begin{subfigure}{0.4\textwidth}
        \caption{Binarized}
        \includegraphics[width=\linewidth]{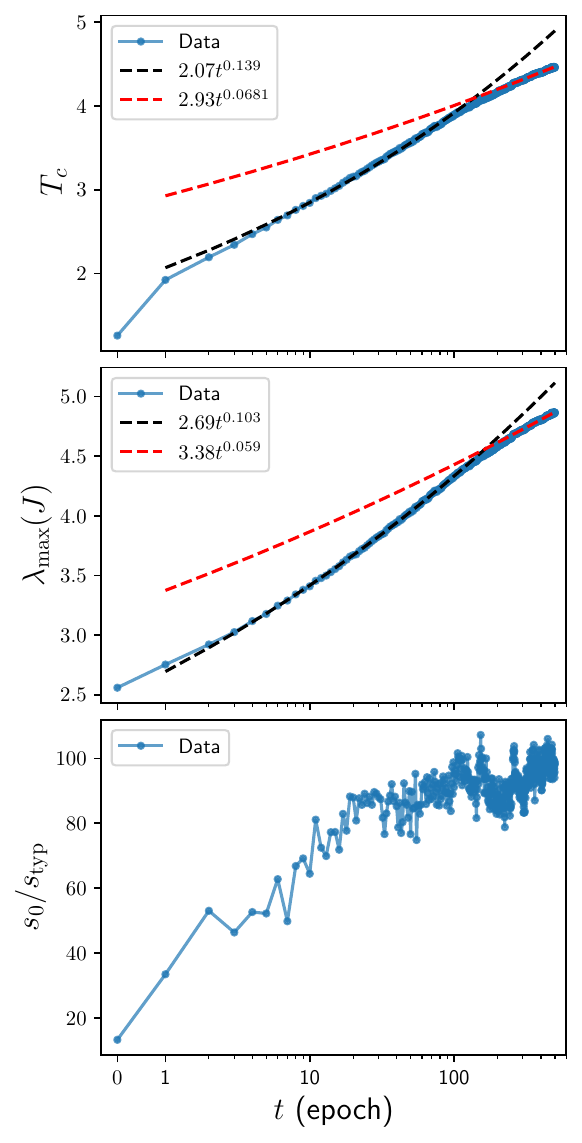}
    \end{subfigure}
    \begin{subfigure}{0.4\linewidth}
        \caption{Standard}
        \includegraphics[width=\linewidth]{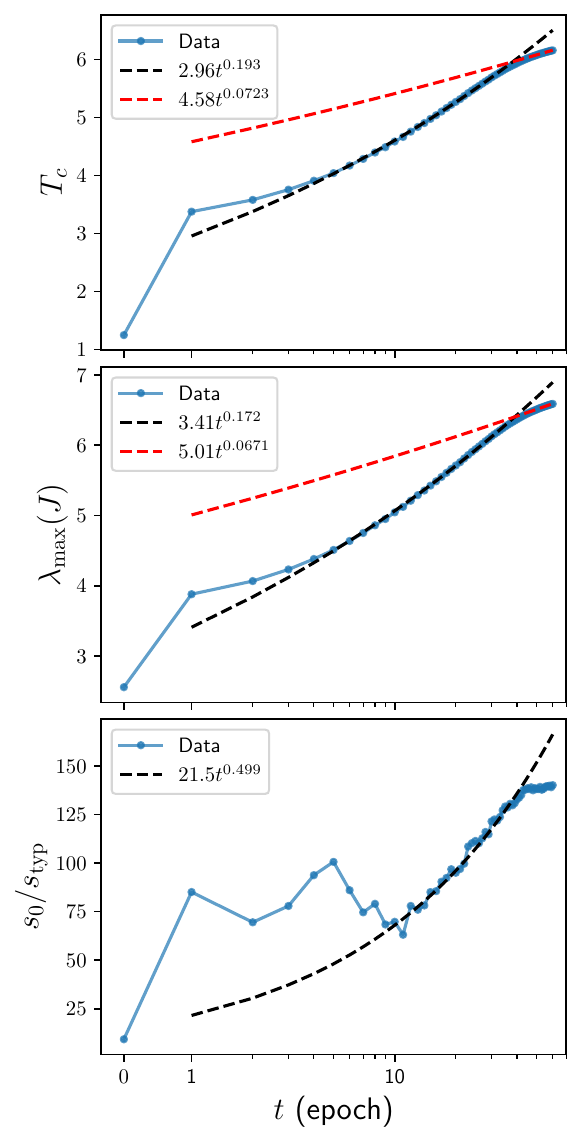}
    \end{subfigure}
    \caption{From top to bottom, the melting temperature, the largest eigenvalue of the bond matrix $J$, and the normalized first level spacing of $I_N-M$ as training progresses. The learning rate for these neural networks is $10^{-4}$. The horizontal axes are logarithmically scaled except for a linear scaling on the interval [0,1].}
    \label{fig:fast_dyn}
\end{figure}

We calculate the transition temperature $T_c$ for each epoch in the training process by finding the value of $\beta$ for which the least eigenvalue curve first vanishes. In order to correct somewhat for finite size effects we find the value of $\beta$ for which the least eigenvalue curve first reaches its minimum value before training. In Fig.~\ref{fig:fast_dyn} we see how the critical inverse temperature evolves with training for our two NNs. Before training the critical temperature is near the analytic result obtained in Section~\ref{ssec:analytic}. After some transitory behavior at early training times the critical temperature begins growing as a power law. Eventually the power-law behavior of $T_c$ appears to change to a power-law growth with a smaller exponent. Comparison with Fig.~\ref{fig:err_rate} shows that this change occurs around the same time at which the validation error rates of the NNs plateau. We surmise that this change in exponents is a result of the NNs reaching their capacity for learning generalizable features of the dataset within their training programs.

\subsection{The nature of the melting transition}

\begin{figure}[h!]
    \centering
    \begin{subfigure}{0.4\textwidth}
        \caption{Binarized}
        \includegraphics[width=\linewidth]{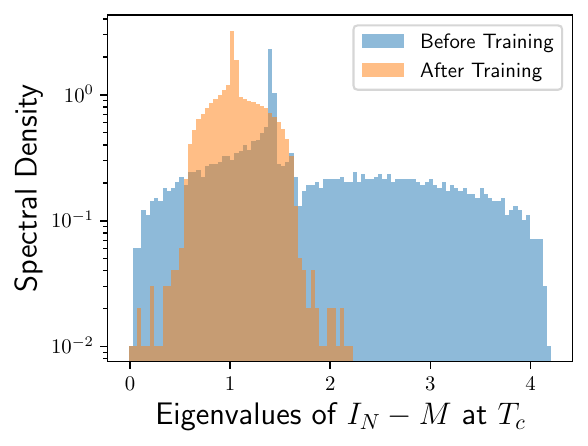}
    \end{subfigure}
    \begin{subfigure}{0.4\textwidth}
        \caption{Standard}
        \includegraphics[width=\linewidth]{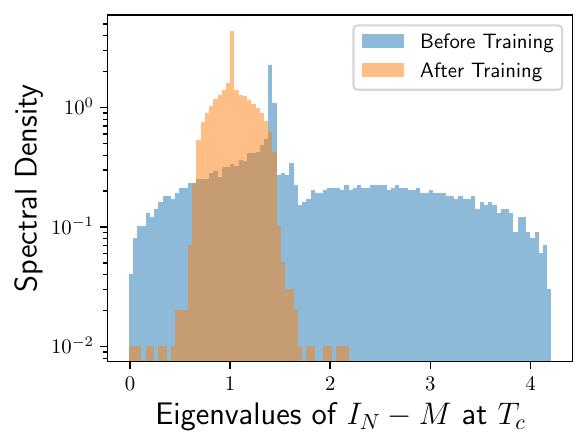}
    \end{subfigure}
    \caption{The spectrum of $I_N-M$ at the transition temperature before and after training. The learning rate for these neural networks is $10^{-4}$. The vertical axes are log scaled.}
    \label{fig:ttemp_spectra}
\end{figure}

In Fig.~\ref{fig:ttemp_spectra} we see the spectrum of $I_N-M$ at $T_c$ before and after training for the two types of NNs at the larger learning rate. As the transition temperature increases with training the spectrum becomes narrower due to the temperature dependence of $M$ seen in Eq.~(\ref{eq:M_gen}). More notably, before training the bulk of the spectrum extends all the way to 0 where it has a hard cutoff. This is typical of a glass transition where an extensive number of TAP solutions appear at the transition~\cite{bray1980}. After training the spectrum has a small number of eigenvalues that have split off of the bulk and form tails, leading to a much lower spectral density at 0. This suggests that the glassy transition away from the paramagnetic phase that exists before training evolves into a different type of transition as training progresses.

In order to further probe this change in the nature of the transition, we track the ratio of the spacing between the smallest eigenvalues $s_0$ to the typical level spacing of the bulk $s_\text{typ}$, which we define as the median level spacing. We refer to this ratio as the normalized first level spacing. The evolution of this quantity is shown in Fig.~\ref{fig:fast_dyn} for the larger learning rate. For both types of NN we observe a large jump in the early training epochs, corresponding with the large increase in $T_c$ at these training times. The standard NN then moves into a period of power-law growth for a time before plateauing. The binarized NN behaves similarly, although no power-law growth for some period of the training is clear.

\begin{figure}[h!]
    \centering
    \begin{subfigure}{0.4\textwidth}
        \caption{Binarized}
        \includegraphics[width=\linewidth]{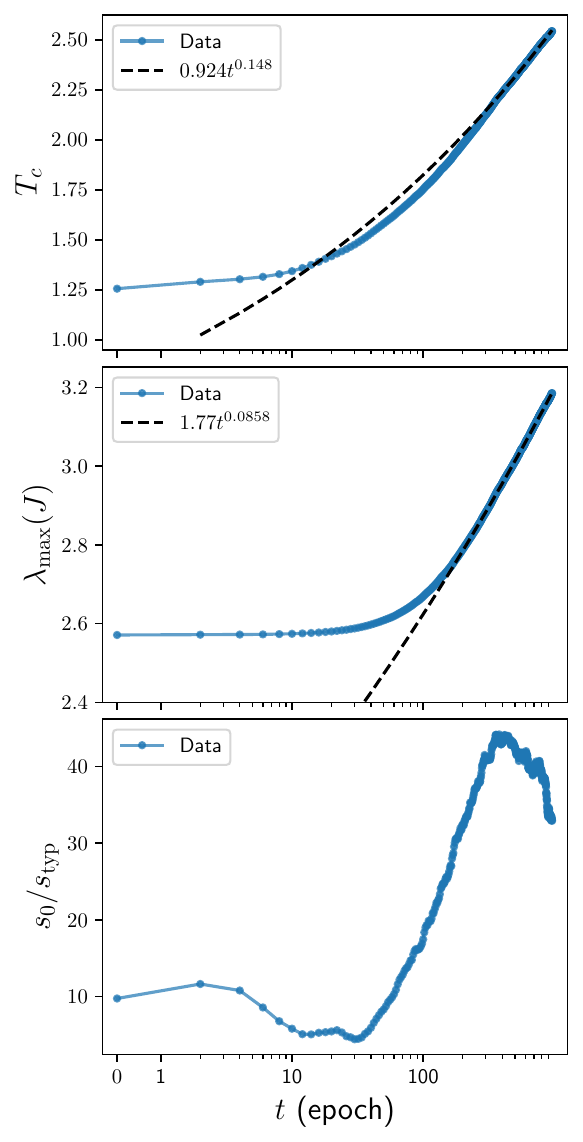}
    \end{subfigure}
    \begin{subfigure}{0.4\textwidth}
        \caption{Standard}
        \includegraphics[width=\linewidth]{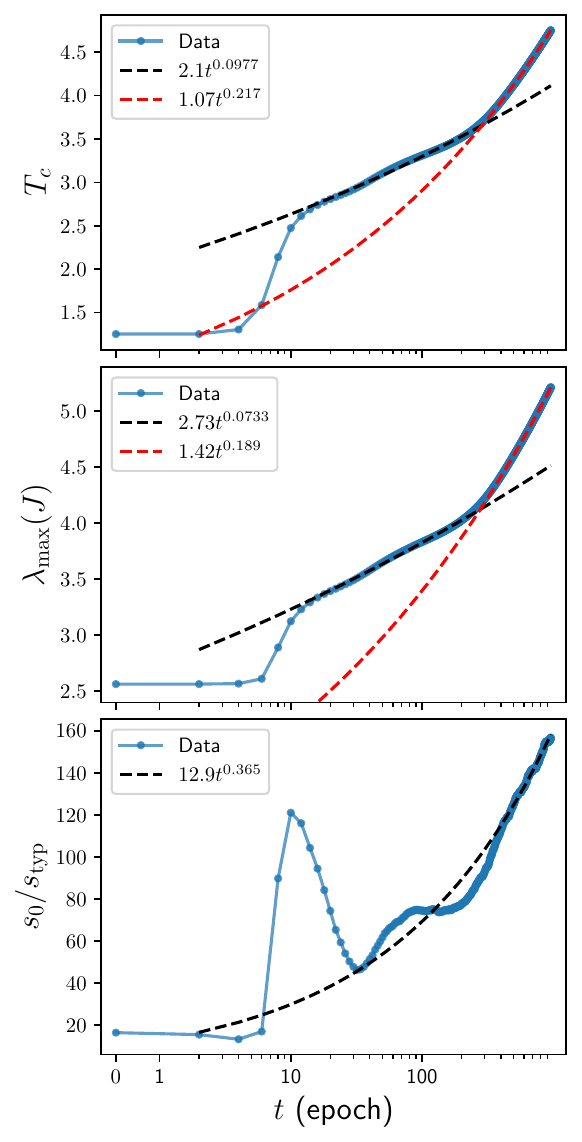}
    \end{subfigure}
    \caption{From top to bottom, the melting temperature, the largest eigenvalue of the bond matrix $J$, and the normalized first level spacing of $I_N-M$ as training progresses. The learning rate for these neural networks is $10^{-6}$. The horizontal axes are logarithmically scaled except for a linear scaling on the interval [0,1].}
    \label{fig:slow_dyn}
\end{figure}

Since this large jump happens very quickly after training begins, it is difficult to determine the nature of this transition. We find that we can delay the onset of this transition by decreasing the learning rate of our training procedure. Fig~\ref{fig:slow_dyn} shows the transition temperature for the two NNs when the learning rate is 100 times smaller than previously. We observe that $T_c$ now stays relatively constant for several epochs before growing. Fig~\ref{fig:slow_dyn} also shows the normalized first level spacing for this reduced learning rate. We observe jumps in this quantity at roughly the same times that $T_c$ begins to grow in the two NNs.

Our interpretation of these results is thus. For both NN types the training procedure, which determines the interactions between spins in the multi-layer spin systems, causes eigenvalues to split off from the bulk of the spectrum of $M$. This causes the critical temperature to grow and greatly reduces the number of TAP solutions which appear at the transition away from the paramagnetic phase, changing the nature of the transition. Instead of a glassy transition at which an extensive number of TAP solutions appear, a single $\mathbb Z_2$ symmetry broken solution will appear. This new phase has some hidden order determined by the training data and procedure. Schematically the phase diagram for spin systems with interactions determined by training will look as in Fig.~\ref{fig:phase_diagram}.

Note that our approach does not probe temperatures significantly less than the melting temperature. As temperature decreases it is possible for a larger number of TAP solutions to appear in the hidden order phase. However, if the largest eigenvalue of $M$ is greatly isolated from the rest of the spectrum it is likely that there will only be a single TAP solution for lower temperatures, potentially all the way down to $T=0$ as in a ferromagnet. Such a situation might arise from overtraining the NN.

\subsection{Evolution of the bond matrix spectrum}

The transition temperature away from the paramagnetic state is determined by the largest eigenvalue of $M$. We see from Eqs.~(\ref{eq:M_block})-(\ref{eq:R}) that for high temperatures the largest eigenvalue of $M$ behaves as $\beta\lambda_{\max}(J)$, while for small temperatures it behaves as $-\beta^2\lambda_{\min}(R)$. For the cases we have studied in this work, the transition temperature is within the crossover of these two regimes, making both $\lambda_{\max}(J)$ and $\lambda_{\min}(R)$ quantities of interest. We observed that $\lambda_{\min}(R)$ hardly changed throughout the entire training process. In no case that we examined did it change by more than 1.2\% of its original value. For this reason we can understand the training dynamics of the multi-layer spin system primarily through the evolution of the spectrum of the bond matrix $J$, in particular its maximum eigenvalue.

\begin{figure}[h!]
    \centering
    \begin{subfigure}{0.4\linewidth}
        \caption{Binarized}
         \includegraphics[width=\linewidth]{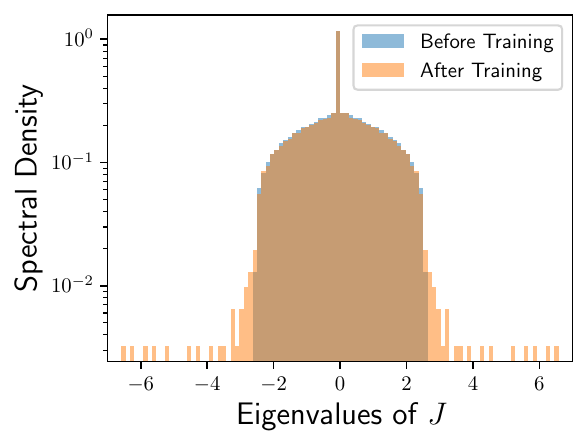}       
    \end{subfigure}
    \begin{subfigure}{0.4\linewidth}
        \caption{Standard}
        \includegraphics[width=\linewidth]{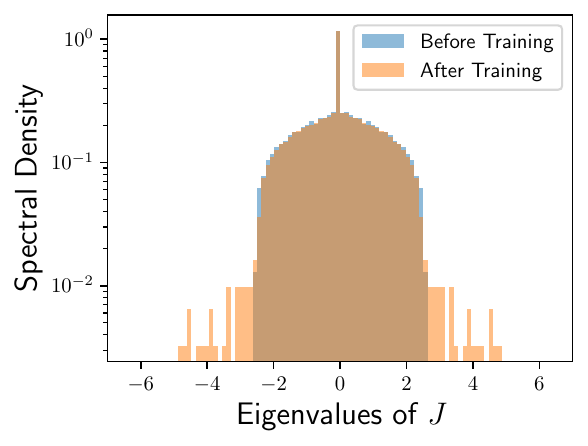}
    \end{subfigure}
    \caption{The spectral density of the eigenvalues of the bond matrix $J$ before and after training. The learning rate for these neural networks is $10^{-4}$. The vertical axes are log scaled. (Left) Binarized neural networks. (Right) Standard neural networks.}
    \label{fig:spectrum_hist_rich}
\end{figure}

Fig.~\ref{fig:spectrum_hist_rich} shows the distribution of the eigenvalues of the bond matrix before and after training for the larger learning rate. We see that the bulk of the spectrum maintains the same distribution but a small number of eigenvalues separate from the bulk and form tails. This phenomenon is likely due to the same processes that have an analogous effect on the spectra of the weight matrices connecting adjacent layers, as studied in other works~\cite{martin2021,martin2021_predicting,matthias2022,wang2024}. It has been noted that, as effective training progresses, the spectra of these weight matrices develop tails outside the predictions of Marčenko-Pastur (MP) random matrix theory~\cite{marčenko1967}. Truncated power-law fits to these tails have been used to assess the quality of models without any knowledge of the training or testing data~\cite{martin2021_predicting}.

The formation of these tails has been connected to rich learning~\cite{matthias2022}, as opposed to lazy learning~\cite{chizat2019,du2018,li2018,du2019,allen-zhu2019,zou2019}. When the NN is in the lazy learning regime the weights change very little and the model performs approximately as its linearization about its initial weights. In this regime the spectra of the weight matrices closely follow the MP prediction with no tails. The appearance of the tails signals that the model is in the rich learning regime which performs significantly better than its linearization, barring overtraining.

In Figs.~\ref{fig:fast_dyn} and \ref{fig:slow_dyn} we see the evolution of the largest eigenvalue of $J$ for the four models we have trained. For the larger learning rate we see that the maximum eigenvalue immediately starts growing and soon behaves as a power-law. This is an indication that the NN has very quickly moved into the rich learning regime. Eventually the largest eigenvalue starts growing as a different power law with a smaller exponent. The time of these changes correspond to analogous changes in behavior in the transition temperature and the normalized first level spacing of $I_N-M$, as seen in Fig.~\ref{fig:fast_dyn}. For the smaller learning rate we see that the maximum eigenvalue initially stays constant, but then quickly moves to a power-law growth. This indicates that the NNs remain in the lazy learning regime for a period of training time before moving to the rich learning regime. Again the changes in behavior of the largest eigenvalue of $M$ occur at the same times as the changes in behavior of the transition temperature and the normalized first level spacing of $I_N-M$, as seen in Fig.~\ref{fig:slow_dyn}.


\section{Conclusion}\label{sec:conclusion}

In this work we have explored a correspondence between NNs and statistical mechanical Ising spin models. We have done this by training two different types of NNs: a standard NN with continuous activations and a partially binarized NN with activations only able to take the values $\pm 1$. At each step in the training process the resulting weight matrices are used to define a bond matrix between spins in a multi-layer spin system analogous to the NN architecture.

We found analytically that, before any training occurs, meaning the weights are all independently and randomly distributed, the multi-layer spin system has a critical temperature separating the ergodic paramagnetic phase at high temperatures from the spin glass phase at low temperatures. Using the TAP equations we numerically determined the transition temperature of the resulting spin system throughout the training. We found that, as training progresses, the transition temperature increases monotonically, exhibiting power-law growth for a time. Additionally, the spin glass transition at which an extensive number of TAP solutions appear is quickly destroyed in favor of a transition to a phase with a single nontrivial TAP solution determined by the training procedure. This suggests a useful unifying paradigm on the training process: training an NN effectively selects and strengthens a small number of symmetry broken states of the multi-layer spin model, one of which is dominant at the transition.

We note that our approach holds only in the regime in which the TAP equations may be linearized, meaning we do not probe temperatures much lower than the melting temperature. Although numerically determining the number of TAP solutions in this low temperature regime is difficult, it may be possible to analytically calculate the average number of TAP solutions in this regime for bond matrices drawn from a random matrix ensemble whose spectra have tails similar to those we have observed after training. Future work may explore this possibility. It also remains to determine how well our results hold for different architectures, datasets, tasks, etc.

In this study we have used the weights resulting from training an NN as bonds in a classical multi-layer spin system. An interesting alternative would be to consider instead a quantum spin system and observe what differences may arise in comparison to the classical case. Such an examination may shed light on the potential of quantum machine learning~\cite{schuld2014,biamonte2017} and would be directly relevant to recent experimental advances in analogue neuromorphic computing in cavity-QED and Rydberg systems~\cite{marsh2021,kroeze2023,marsh2024}. We plan to explore this avenue in a future work.

\section*{Acknowledgements}
V.G. is grateful to Ben Lev for useful discussions. This research was sponsored by the Army Research Office under Grant Number W911NF-23-1-0241, the National Science Foundation under Grant No. DMR-203715, and  the NSF QLCI grant OMA-2120757. M.W. acknowledges support from the Joint Quantum Institute.
\printbibliography

\end{document}